# Convolution Forgetting Curve Model for Repeated Learning

Yanlu Xie[*1], Yue Chen[1], Man Li[1]

1College of Information Science, Beijing Language and Culture University, Beijing, China

*xieyanlu@blcu.edu.cn*

**ABSTRACT:** Most of mathematic forgetting curve models fit well with the forgetting data under the learning condition of one time rather than repeated. In the paper, a convolution model of forgetting curve is proposed to simulate the memory process during learning. In this model, the memory ability (i.e. the central procedure in the working memory model) and learning material (i.e. the input in the working memory model) is regarded as the system function and the input function, respectively. The status of forgetting (i.e. the output in the working memory model) is regarded as output function or the convolution result of the memory ability and learning material. The model is applied to simulate the forgetting curves in different situations. The results show that the model is able to simulate the forgetting curves not only in one time learning condition but also in multi-times condition. The model is further verified in the experiments of Mandarin tone learning for Japanese learners. And the predicted curve fits well on the test points.

## 1.Introduction

Since Ebbinghaus proposed the eminent 'forgetting curve' to quantitatively describe the procedure of memory [1], researchers have made great effort to find out the exactly form of the models for forgetting curve. The problem is still being considered as 'central theoretical importance' [2]. However, there are few well-accepted curve models yet. The forgetting curve describes basic, lawful, and robust memory phenomena, simulates and predicts the memory tasks[3][4][5].

Most of the shapes of the previously proposed forgetting curves are more or less a curvilinear function of time [6][7][8][9][10]. Wixted compared a few typical functions, i.e. power, linear, exponential, exponential-power, hyperbolic and logarithmic functions. His and other's experiments showed that power function outperformed the others and described the courses of forgetting best [6][11],which was also confirmed in Murre's replication experiment [12]. In Wixted's experiments, the forgetting curves derived from short-term recall of words, long-term recognition of faces in humans, and delayed-matching-to-sample in pigeons. The best function may be different when conditions change.

Thus more complicated and applicable models were proposed. For example connectionist models of biologically are reasonably used to predict exponential forgetting curves and outperform most of two-parameter forgetting-curve functions mentioned by Wixted[3]. Also hierarchical Bayesian Models were fitted in to handle individual participant data[13] .Besides, some researchers have combined psycholinguistics-inspired memory models with modern machine learning techniques and proposed half-life regression (HLR) as a trainable interval repetition algorithm [14]. These models were able to avoid distortions by averaging over participants. However, the forms of the functions became more and more complex and the parameters in the function were difficult to be determined [15]. Some researchers claimed that a two-parameter function can explain more than 90% of the variance in the psychology memory [16]. Actually, in terms of one-time training condition, all of the typical functions mentioned by Wixted could fit the forgetting curve well.

Wixted proposed the basic form of exponential function in forgetting [6]. Here, the function is written as

$$f(t) = a_1 \exp(-a_2 t) + a_3 \quad\quad （1）（t \text{ in minutes}）$$

Parameters a1 a2 a3 represent the personal intrinsic characteristic of the learner, where a1 is the coefficient representing the initial memory strength, a2 represents the strength of the rate of forgetting as to time, a3 means long time retain, i.e. the learned material.
Formula (1) is a little different to the original function in parameter a3. It is because that in view of long term memory,

some material will turn into long term memory in learner's mind. The original exponential function could not reflect the phenomenon and the revised formula are proved to fit the remaining memory well. In Yifeiyi's experiments, with parameters a1=0.3796, a2=1.475, a3=0.211, formula (1) simulated the forgetting curve of Japanese language learning well [17].

In Rubin's three continuous cued recalls experiments[18], the basic forgetting function is described as

$$f(t) = a_1 \exp(-t/T_1) + a_2 \exp(-t/T_2) + a_3 \quad (2)$$

T1 is fixed to 1.15 and T2 is fixed to 27.55. The formula is similar to formula (1) and simulates the recall data well.

The forgetting experiments are usually carried out several minutes or some days later after learning one time. However a few experiments were carried out to simulate the variation of forgetting curves in multi-training conditions. In the situation learners may be trained with some material every day or every few days repeatedly to remember the material. And the learners are tested in the period to see the memory retention. It is a more general procedure of learning like language learning during which the single forgetting curve could not model the data directly [19][20].

There exists sophisticated relation between the memory and the forgetting curves. The forgetting curves are impacted by the pattern of the memory. In other words, the forgetting curves could be viewed as the outer behaviors of the memory. Most of memory models proposed that memory consists three important stages, i.e. input, output and central working memory. Baddeley introduced the concept of working memory[21][22][23][24]. Working memory is crucial to the higher cognitive functions, such as language learning and comprehension, mathematical abilities and reasoning[25]. It also affects the elements in the long term memory[22][26]. As shown in Figure 1, working memory is supposed to be a storage area while learning material is been orchestrated.

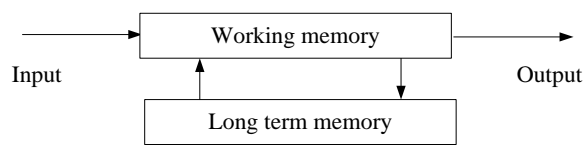

**Figure.1** Working memory systems for language learning [27]

From the Figure 1, most of the two-parameter function for forgetting curves mentioned by Wixted cannot explain the working memory procedure directly. Firstly, the two-parameter function cannot explain how the long term memory is formed. In the simulation curves, the memory retention will be zero with time passing by. Then the long term memory could not be set up. More parameters must be added to represent the learned material. Secondly the function cannot explain multi-times learning. In this kind of learning, working memory interacts with long term memory.

In most of working memory model, the learning procedure can be qualitatively evaluated [28][29]. Under some circumstances, we also wonder the precise memory result in learning. The results are useful in assessing the learning method and the individual personality of the learner. The adjusted forgetting curve functions may be used to quantitatively evaluate the procedure.

A typical example of memory process of Figure 1 is language learning where repeated training is applied frequently in order to get the learner's attention and establish long term memory. This process cannot be directly modeled as the two parameters forgetting curves are not accommodated to multi-training conditions.

For example, experiments conducted by the researchers in the University of Waterloo[30] showed that within 24 hours of getting the information spend 10 minutes reviewing and the forgetting curve can be raised almost to 100%. A week later (day 7), it only takes 5 minutes to "reactivate" the same material, and again raise the curve. This shows the function repeated learning, as shown in Figure 2. The left curve is the remember variation with time from one time learning to forgetting. The right curve shows the effect of multi-times learning, the curve could arrive at 100% after the repeated learning.

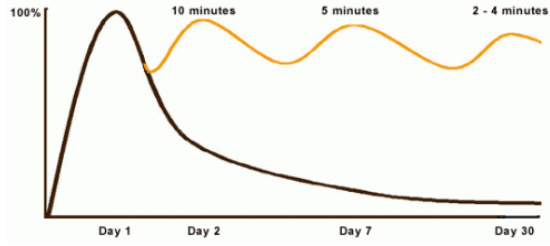

**Figure 2.** Forgetting curve from University of Waterloo

As to language learning, for the purposes of memory formatting, training methods always depend on a number of voice loops to stimulate the central executive as illustrated in memory model of Figure 1, so that the voice-based short-term memory can be turned into long-term memory. The variation of forgetting curve is more similar to that of the right curve in Figure 2. On the other hand, most of the mathematic models are devoted to simulating left curve. Such models can be a good description of a single memory with time attenuation. And current models face much difficulty in simulating right curve. In the following section, parameterized convolution forgetting model is proposed to simulate the left and right curves in Figure 2.

## 2. Results

### 2.1. Convolution Forgetting Curve Model

From the Figure 2, it can be found that through repeated learning or remembering, the memory in the brain is similar to a decrease exponent function, and studying once, the information left in the brain will have a step change.

From the original memory model illustrated in Figure 1, long-term memory conformation is the result of interaction of input and the central executive in the working memory. In consideration of the relationship between stimulation (study) and memory, it is alike interaction of signal and system in circuit theory [31].

Thus a convolution model for memory is presented. The material for study is regarded as a stimulation signal. The central executive is supposed to be central system. The result of convolution is the information left in the brain after studying $n$ times.

Compared with the convolution calculation in mathematics, the interaction of input $f(t)$ and the central executive $h(t)$ is viewed as convolution analysis. Provided that the central executive is linear time-invariant (LTI), the output $y(t)$, i.e., the memory result can be written as

$$y(t)=\int_{-\infty}^{\infty} f(\tau)h(t-\tau)d\tau = f(t)*h(t) \qquad (3)$$

In the procedure of language learning, $f(t)$ is regarded as one's ability of remembering, $h(t)$ is regarded as voice stimulation, $y(t)$ is forgetting curve. The formula is the general format for forgetting curve. In order to simulate the curve specific in learning situation, the function form and its parameters should be determined before.

Considering the complexity of memory, the proposed forgetting curve models are classed into the following three different specific kinds.

*2.1.1. One time learning convolution model (OCM)*

This model describes the classic forgetting curve as in the left curve in Figure 2.

From a biological perspective, people's memory comes from stimulation, and the trace in the cortex of brain made by stimulation could be viewed as an output or a response. The unit impulse function $\delta(t)$ can be assumed as the simplest stimulation. Thus Formula (3) can be written as:

$$y(t)=\delta(t)*h(t)=h(t) \qquad (4)$$

Formula (4) can also be explained in terms of circuit and systems. If the LTI system $h(t)$ has been in a steady state, its response is called zero-state respond. The convolution of unit impulse function and the system is the result of the respond, i.e, the system itself in mathematics. This also reflects the characteristic of the system. In another word, the *h(t)* is the forgetting result reflecting the capability of the brain. The respond *y(t)* to continuous, time invariant, linear system is convolution of excitation *f(t)* and the unit impulse function respond *h(t)*.

*h(t)* could be viewed as the basic forgetting function. Taking formula (1) for example, in the situation, *y(t)* is equal to *h(t)* and can be written:

$$y(t) = h(t) = a_1 \exp(-a_2 t) + a_3 \quad (5)$$

Therefore, in the final form, OCM is exactly the same as classic exponential function of forgetting curve. The different between them is that OCM is derived from the point of view of impulse and memory system. Formula (4) can be adapted to any forgetting function by adjusting *h(t)*.

*2.1.2.Repeated learning convolution model (RCM)*

This model describes the right forgetting curve in Figure 2.

If the learners have been stimulated several times with the same material (signal), the impulse function should be presented corresponding times. Thus the Formula (3) can be written as:

$$y(t) = \sum_{n=1}^{N} f(t - T_n) * h(t) \quad (6)$$

Where *n* means the times of stimulation or learning and $T_n$ is the time delay comparing with the first of stimulation at the *n*th time. If the impulse is still impulse signal, the formula is written as:

$$y(t) = \sum_{n=1}^{N} \delta(t - T_n) * h(t)$$
$$= \sum_{n=1}^{N} h(t - T_n) \quad (7)$$

If the memory system is LTI and Formula (1) is also taken as the basic function, then the final form can be written as:

$$y(t) = \sum_{n=1}^{N} a_1 \exp[-a_2(t - T_n)] + Na_3 \quad (8)$$

Formula (8) implies that the repeated impulses lead to accumulating of the memory.

According to the analysis of LTI system and convolution, it is hypothesized that the procedure of remembering is similar to a process of the memory center being stimulated repeatedly. The response of memory is convolution result of the system forgetting curve and the system impulse. The response of repeated training memory is sum of the convolution result of the system forgetting curve and the system impulse. The convolution memory model illustrated in Figure 3 is pretty much similar as the working memory model proposed by Baddeley [19].

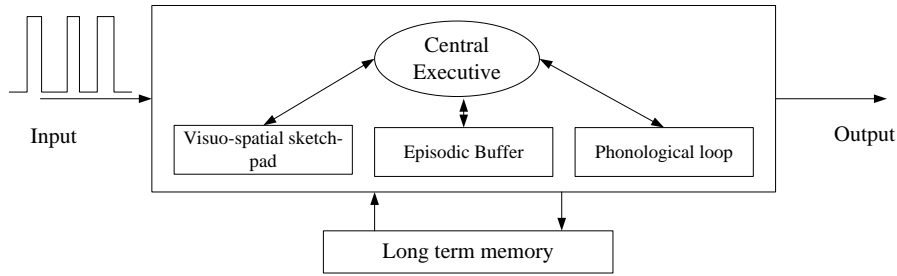

**Figure 3.** Procedure of convolution memory model[21]

In Figure 3, the signal stimulates the central executive system and leaves impression in the buffer. With time passes by, the impression or the short time memory become weak. If the signals are input from visuo-spatial sketchpad repeatedly, i.e. the learning is repeatedly, episodic buffer will accumulate more and more impression. Just like formula (8), the convolution result will be summed up. With phonological loop, the learning material turns into long term memory. Therefore the proposed forgetting curve model is consistent with the working memory model and explains it in mathematics.

Generally the function which has interaction with *f(t)* will be unit impulse function or its shifts. Furthermore the input is not always the impulse function; the material of learning will affect the stimulation input function's format. So the more general input function form is proposed as follows.

*2.1.3. General repeated learning convolution model (GRCM)*

The input in the convolution model can be an arbitrary function other than the impulse function in 3.1 and 3.2. As we known, the procedure of remembering something is time consuming. And the memory of material provided to the learners will not disappear quickly. It may last several seconds or minutes and sometimes even longer. Thus the impulse function should be replaced with a function that can last for longer time.

Also with the result 3.1 and 3.2, the left rising trend in the left and right forgetting curves of Fig.2 could not be simulated. This trend reflects the learning procedure in the curve.

Here a rectangular function is used to replace impulse function in Formula (4) and (7). Formula (1) is also taken as the basic function. Thus the two formulas are rewritten as Formula (9) and (10) respectively.

$$y(t) = [u(t-\tau) - u(t)] * h(t) = [u(t-\tau) - u(t)] * [a_1 \exp(-a_2 t) + a_3] \quad (9)$$

$$= \begin{cases} \frac{a_1}{a_2}[1 - \exp(-a_2 t)] + a_3 t & 0 < t < \tau \\ \frac{a_1}{a_2} \exp(-a_2 t)[\exp(a_2 \tau) - 1] + a_3 \tau & t > \tau \end{cases}$$

$$y(t) = \sum_{n=1}^{N} [u(t - T_n - \tau) - u(t - T_n)] * h(t) = \sum_{n=1}^{N} [u(t - T_n - \tau) - u(t - T_n)] * [a_1 \exp(-a_2 t) + a_3] \quad (10)$$

$$= \begin{cases} \sum_{n=1}^{N} \frac{a_1}{a_2}[1 - \exp(-a_2(t - T_n))] + a_3(t - T_n) & 0 < t < \tau \\ \sum_{n=1}^{N} \frac{a_1}{a_2} \exp(-a_2(t - T_n))[\exp(a_2 \tau) - 1] + a_3 N\tau & t > \tau \end{cases}$$

In the above formula, $u(t)$ refers to step function, $\tau$ is the width of rectangular function, which is viewed as the learning continuous period for one time. In Formula (10), the period is assumed to be equal each time.

With some derivation, formulas (9) and (10) are written as the segment functions, which mean that the curves are divided into two stages. In the first stage, the curves are rising which corresponds with the first part of the segment function. And the learner is continuously learning along with remembered contents of the material increasing. In the second stage, the curves are falling which corresponds with the second part of the segment function. And the learner stops learning with its forgetting curve decline.

Formulas (9) and (10) are both based on the basic formula (1). If the basic forgetting function is replaced with formula (2), similar formulas can also be achieved.

**2.2. Simulation Results**

*2.2.1. Experimental background*

Mandarin tone is used here as the repeated learning material for Japanese learners to test the forgetting curve model. Mandarin is a tonal language, which means different tones express different meanings. It has been found that Japanese learners have great difficulties in Mandarin tonal pronunciation, especially Tone 2 and Tone 3.

In order to help Japanese learners correctly produce tones of Mandarin and improve their learning efficiency, perception training method was applied [32][33]. The method tried to correct learners' bias by repeating learning same material in a short time.

This training is a typical repeated learning procedure, which cannot be described by the traditional model of one time forgetting curve. And it is a good example to verify the proposed convolution memory model.

*2.2.2. Perceptual training experiment*

The learners in the perception training experiment are eleven Japanese students who are all studying Mandarin in China. The main purpose of the perception training is to test the variation of Japanese's aesthesis about Mandarin Tone 2 and Tone 3. The cognitive training includes training module and testing module. They are clarified as follows.

Training module: this module includes two parts i.e. adaptive perceptual training and high variability training [31]. The training material includes 20 and 60 words. The 20 words training (20 trails) and the 60 words training (60 trails) are both carried from day 1 to day 6.

Testing module: The test materials are same as the training. Learners are requested to judge the words' tone in 5 minutes. In the 60 trails situation, the tests are carried on day 1 and day 3. On the two days, the tests are taken before the training. A plus test is taken on day 7 days to verify to remaining memory. In the 20 trails situation, the tests are carried from day 1 to day 6. And on these six days, a test is taken after the training. Since the learners are foreign students, two of them left China in the procedure and only nine learners participated in all experiments of the 60 trails.

The test results of 60 trails and 20 trails are shown in Table 3 and Table 4, respectively. It can be seen that after the perception training, most of the learners' overall tonal level shows a rising trend. Therefore the traditional forgetting model is not suitable to describe this type of situation.

**Table 3.** The probability of recall for the experiments of 60 trails

| Learner \Day | 1 | 3 | 7 |
|---|---|---|---|
| 1 | 0.60 | 0.75 | 0.80 |
| 3 | 0.87 | 0.75 | 0.92 |
| 4 | 0.68 | 0.85 | 0.85 |
| 5 | 0.85 | 0.93 | 0.95 |
| 6 | 0.93 | 0.98 | 0.98 |
| 8 | 0.97 | 0.95 | 0.92 |
| 12 | 0.97 | 1 | 0.98 |
| 13 | 0.87 | 0.98 | 0.98 |
| Avg | 0.843 | 0.899 | 0.941 |

**Table 4.** The probability of recall for the experiments of 20 trails

| Learner \Day | 1 | 2 | 3 | 4 | 5 | 6 |
|---|---|---|---|---|---|---|
| 1 | 0.75 | 0.6 | 0.75 | 0.95 | 0.9 | 0.8 |
| 2 | 0.9 | 0.9 | 1 | 1 | 1 | 0.95 |
| 3 | 0.95 | 0.65 | 0.95 | 0.9 | 1 | 1 |
| 4 | 0.65 | 0.85 | 0.9 | 0.85 | 0.9 | 0.9 |
| 5 | 0.85 | 0.95 | 0.75 | 0.65 | 0.85 | 1 |
| 6 | 0.8 | 1 | 0.95 | 1 | 1 | 1 |
| 7 | 0.9 | 0.95 | 0.75 | 0.85 | 0.85 | 0.95 |
| 8 | 0.85 | 0.8 | 0.95 | 1 | 0.95 | 0.95 |
| 9 | 0.9 | 0.95 | 0.85 | 0.85 | 0.9 | 0.9 |
| 10 | 0.85 | 0.9 | 1 | 0.95 | 0.95 | 1 |
| 11 | 0.9 | 0.9 | 0.85 | 0.9 | 0.9 | 0.9 |
| Avg | 0.85 | 0.86 | 0.88 | 0.9 | 0.93 | 0.94 |

**2.3. Simulation for the experiments of 60 trails**

From Table 3 and 4, it can be seen that each student's memory level are not the same. It is due to the difference of student's pronunciation basis and the student's own learning ability. The latter is corresponding to the unit impulse response $h(t)$ proposed in section 3 in the convolution forgetting model.

According to Formula (8), the main different of $h(t)$ is the parameters $a_1$, $a_2$, $a_3$ for each person, where $a_2$ represents the decline speed of the forgetting curve. The bigger $a_2$ is, the faster the curve declines; $a_1$ determines the starting point of the curve, and $a_3$ means the memory result when time extends to infinity. The merit of the forgetting curve model is also consistent with the long term memory in the Baddeley's memory model theory. Parameter $a_3$ represents the long term memory which is reserved in the long time. If the basic function in Formula (8) is replaced with Formula (2), $a_1$ plus $a_2$ determine the starting point of the curves, and $a_3$ also represents the long term memory.

Table 5 shows the calculated $a_1$, $a_2$ and $a_3$ of 60 trails tests results. The average forgetting curves with two different formulas are shown in Fig. 7.

**Table 5.** The calculated $a_1$, $a_2$, $a_3$ with 60 trails tests result (Averaged)

| formula | MSE | MSE of day 1 and3 | MSE of day 7 | r2 | a1 | a2 | a3 |
| --- | --- | --- | --- | --- | --- | --- | --- |
| 1 | 0.001 | 0.001 | 0.003 | 1 | 0.05 | 0.13 | 0.14 |
| 2 | 0.004 | 0 | 0 | 0.984 | 0.28 | 0 | 0.14 |

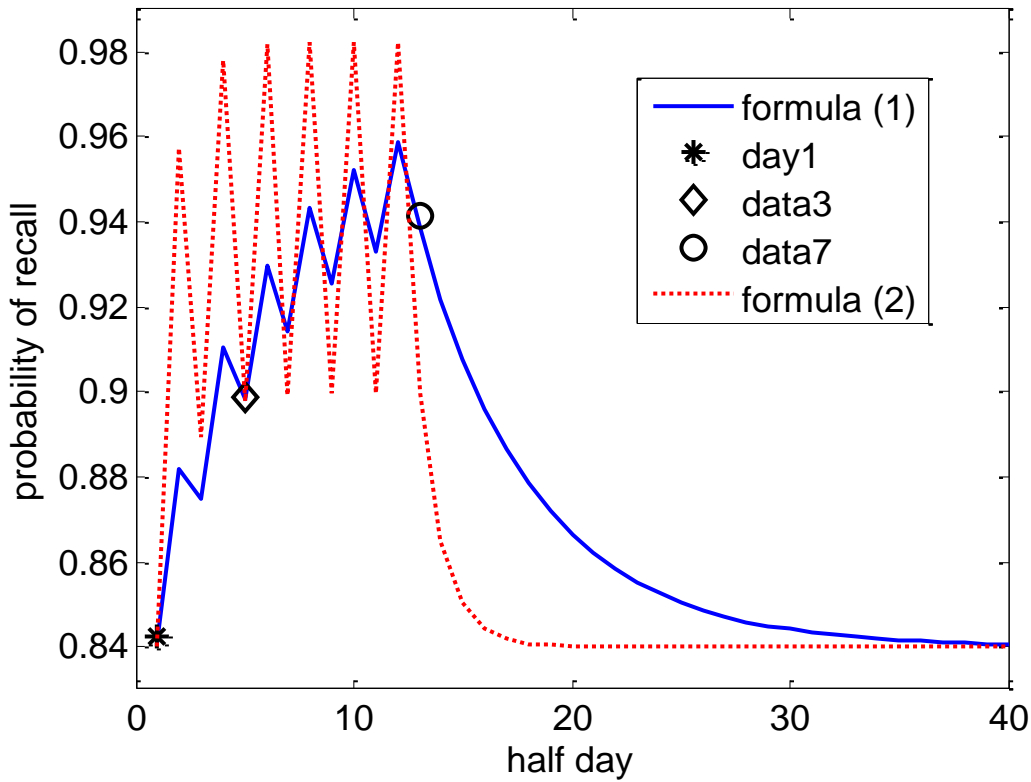

**Figure 7.** The forgetting curves of convolution model (average)

The table and the figure show that the proposed convolution model could fit the average forgetting data well. The r-square and the MSE both demonstrate the effectiveness of the curve.

Figure 7 is alike the right curve in Figure 2 in the shape. It describes forgetting curves under the six times learning situation. The six maximum points in each curve in Figure 7 reflect the actual six learning time points. At the time the memory reach the local optimal states. And with time passing by, the curve declines. Therefore, it is possible to analyze the procedure of memory using convolution memory model in the shape.

The calculated $a_1$ and $a_2$ of each formula show the two different forgetting functions. This means the forgetting curves are different. From Figure 7 it can be seen that the two curves are not coincide except for several point. It may be due to the lack of test data. If the forgetting status of day 2, 4, 5 and 6 are all recorded, the two curves may be calculated more precisely. However $a_3$ is the same. It suggests that the long term memory of the two formulas is same and the forgetting curves in Figure 7 also prove it. The forgetting status of day 7 is predicted well by both of the two curves. The MSE of day 7 is 0.003 and 0 respectively. It shows the ability to predict forgetting status of the proposed model.

**2.4 Simulation for 20 trails experiments**

In the above experiment, parameters $a_1$, $a_2$, $a_3$ are calculated from three days' forgetting result of 60 trails. The learning procedure and the prediction capability of the proposed model can be verified by the test result of 20 trails. Since the 20 trails are tested after six days' learning, the three parameters can be calculated from results of only one day or combination results of some six days or even all results of the six days.

Table 6 shows the MSE and $r^2$ calculated by the result of different number of day of 20 trails. It can be found that the more results are used, the MSE is smaller. It suggests that the forgetting curves fit the data better. Also when more than three days' results are used, the r-square exceeds 0.9, it means the correlation of the curves and the data.

**Table 6.** The MSE and $r^2$ with 20 trails tests result

| Train day | formula | MSE of all day | MSE of train day | MSE of predict day | r2 |
|---|---|---|---|---|---|
| 1 | 1 | 0.023 | 0 | 0.028 | 0.563 |
|   | 2 | 0.018 | 0 | 0.022 | 0.906 |
| 1 and 2 | 1 | 0.018 | 0 | 0.027 | 0.570 |
|   | 2 | 0.015 | 0 | 0.022 | 0.923 |
| 1,2 and 3 | 1 | 0.003 | 0.002 | 0.006 | 0.992 |
|   | 2 | 0.007 | 0.003 | 0.014 | 0.931 |
| 1,2,3 and 4 | 1 | 0.003 | 0.002 | 0.008 | 0.992 |
|   | 2 | 0.004 | 0.002 | 0.010 | 0.966 |
| 1,2,3,4 and 5 | 1 | 0.003 | 0.003 | 0.010 | 0.992 |
|   | 2 | 0.002 | 0.002 | 0.001 | 0.982 |
| 1, 2, 3, 4,5 and 6 | 1 | 0.002 | 0.002 | 0 | 0.977 |
|   | 2 | 0.002 | 0.002 | 0 | 0.982 |

Figure 8 and 9 show the forgetting curves calculated with four days and six days' results respectively. Since the test is carried after the training, the black points which represent the forgetting status of the test time are close to the maximum points of the curve. In order to reflect the phenomenon in the curve, the procedure of calculating $a_1$ $a_2$ and $a_3$ in the model is adjusted in minimizing the mean squared error between the experimental $d'$ and the predicted $d$ in the learning point.

From Figure 8, it is found that the difference of the two curves is significant. Maybe the results used to evaluate the parameters are not enough. And the two curves both fit first four days' points well. However the predicted data of day 5 and day 6 does not coincident with real data. It is also reflected in MSE of predict day in table 6.

In Figure 9, the two curves are nearly the same. It may suggest that with enough evaluated data, formula (1) and formula (2) can derive similar forgetting curves. Also the two curves both fit all of the six points well.

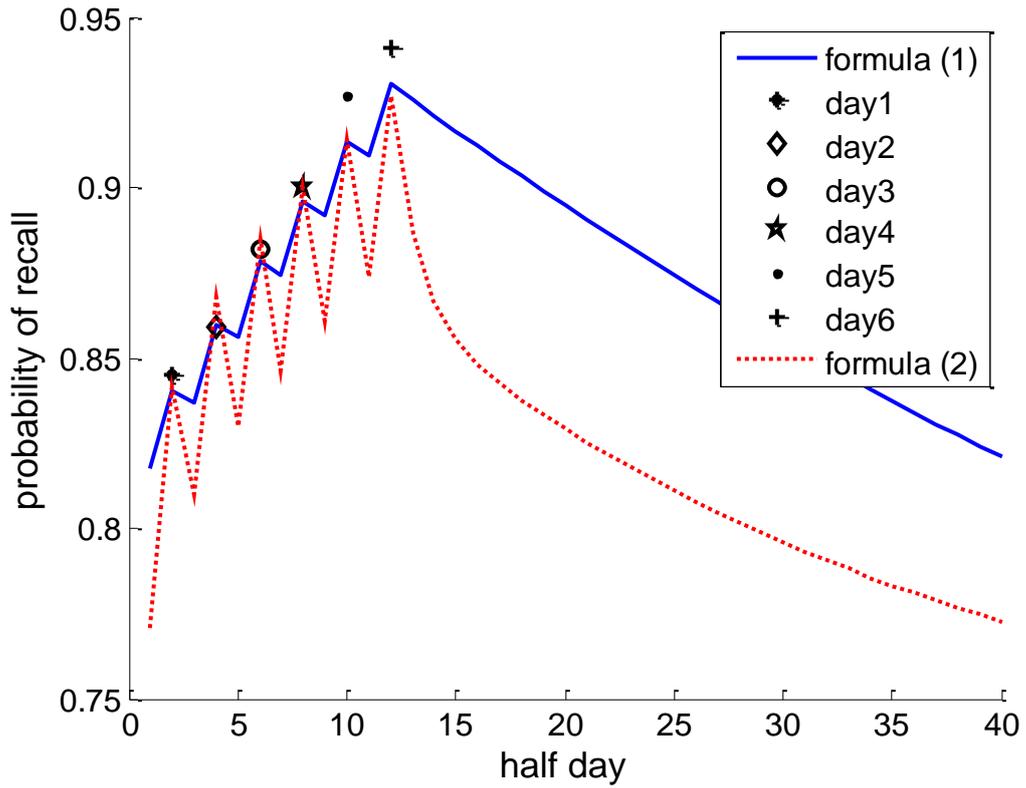

**Fig. 8** The forgetting curves of convolution model (calculated with 4 days' results)

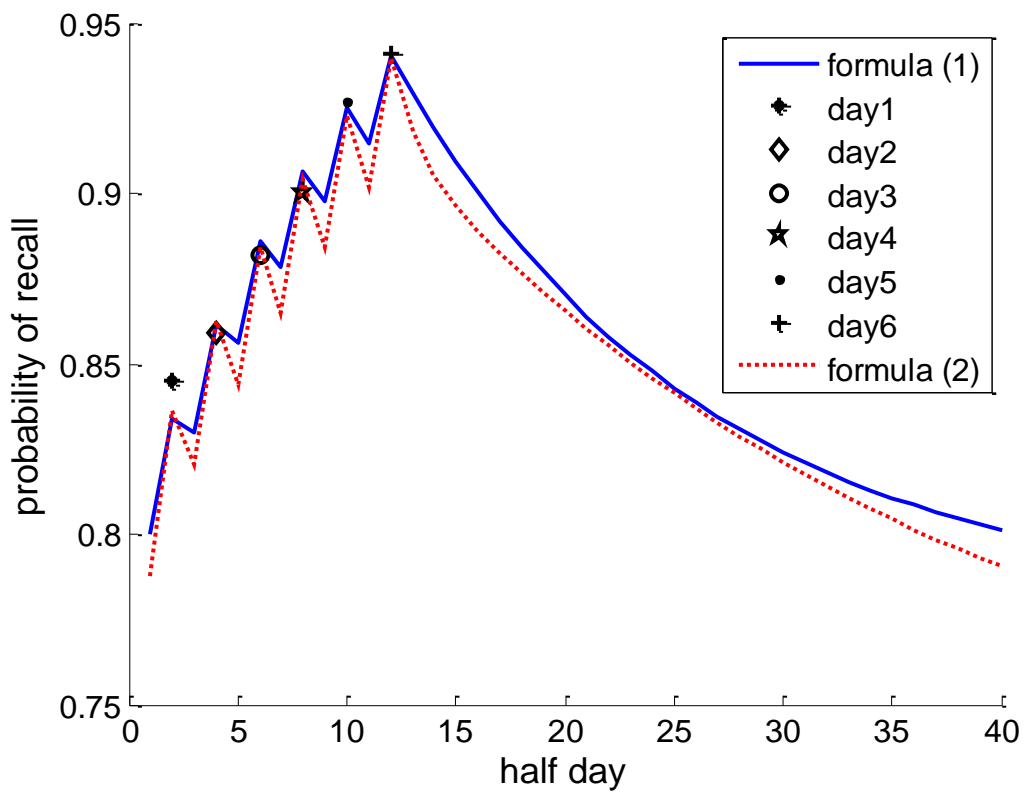

**Figure 9.** The forgetting curves of convolution model (calculated with 6 days' results)

**3. Discussion**

### 3.1. Analysis of the simulation results

Mandarin perceptual training is a typical repeated learning procedure. The above experiments marked the particular learning or forgetting points and simulated the forgetting curve with the proposed model.

From the result of the experiment, it can be seen that the forgetting procedure largely corresponds to the proposed model. The forgetting curves are similar in the appearance. If the training data for $a_1$, $a_2$, $a_3$ is enough (three days or above), the MES and $r^2$ of the propose models are satisfactory. The amount of the training data may affect the fit result. In the 60 trails experiment, the two curves based formulas (1) and (2) respectively. The curves only overlap in several forgetting points. So do the curves calculated from 4 days' results from the experiments of 20 trails. However when calculated with 6 days' results from the experiments of 20 trails, the two curves are nearly the same.

Maybe the details of the curves are different for each learner, because every learner's capacity and starting point of study are different. So in our plots, only the average results are shown.

### 3.2. Discussion

In this paper, a general convolution forgetting curve model is proposed. The model is compactable to the working memory theory, and the traditional forgetting curve can be viewed a particular case of the proposed model. With this model, we analyzed the result of Mandarin perception training. The experimental data is according to the hypothesis of the model both in the 60 and 20 trails.

Via the model, the condition of language study and memory can be analyzed to fit the amount of memory at special time. So that reasonable stimulation point of study can be set up. It helps to provide theory evidence for designing reasonable stimulation points in language learning.

This model also provides a certain basis to design better teaching methods. As to arriving at some memory condition, the learners can be design to learn some material in some time and in some times, the desired forgetting percent can be calculated by the proposed model. With the proposed model, the memory condition at specific time point can be predicted using only a few test data. Thus the two methods can be compared in terms of the amount of memory in the specific time.

Moreover in the future study, the convolution forgetting curve model should be improved in follow aspects.

a) In the above experiment, only the inherent factors are considered in the forgetting curve. External factors may also influence the curve, such as teaching skill, stimulus intensity. The proposed model can also include such factors. In the model, those can be regarded as the strength of the stimulus. Through adjusting the input impulse function coefficient, the situation can be simulated.

b) In the proposed model based on the two basic formulas, parameter $a_3$ refers to the remaining memory in the brain. In the simulation curves and the experiments, the memory amount becomes small with time. It can be derivate that the memory amount will be zero after very long time. It is a little conflict with the long term memory theory. In the future research, parameter $a_3$ may be adjusted as a function of number of times, otherwise a constant. It may increase with the learning time. Thus even the time is very long, there also some impression is left and the long term memory in formulated.

c) In addition, in this paper, the language learners only were trained with Mandarin tones. Also the long time forgetting condition was not tested and verified in the experiments due to limited time. In the future, different training material will be used to further verify the accuracy of the model.

### 4. Methods

As to the simulation experiments, the learners are trained with 60 or 20 trails one time in each six days. Provided that one day is divided into two time interval for learning and forgetting, according to the three training time point, the six impulse functions in Formula (7) can be written as:

$\delta(t-1) = [0\ 1]$

$\delta(t-3) = [0\ 0\ 0\ 1]$

$\delta(t-5) = [0\ 0\ 0\ 0\ 0\ 1]$

$\delta(t-7) = [0\ 0\ 0\ 0\ 0\ 0\ 0\ 1]$

$\delta(t-9) = [0\ 0\ 0\ 0\ 0\ 0\ 0\ 0\ 0\ 1]$

$\delta(t-11) = [0\ 0\ 0\ 0\ 0\ 0\ 0\ 0\ 0\ 0\ 0\ 1]$

Under six stimulations and according to convolution memory model, Function (7) can be written as:

$$\begin{aligned} y(t) &= \delta(t-1)*h(t) + \delta(t-3)*h(t) + \delta(t-5)*h(t) + \delta(t-7)*h(t) \\ &+ \delta(t-9)*h(t) + \delta(t-11)*h(t) \\ &= h(t-1) + h(t-3) + h(t-5) + h(t-7) + h(t-9) + h(t-11) \end{aligned} \quad (11)$$

The formula means that the result of the six stimulations is the sum of results of six single simulations with different time delays. The stimulation interval is half day.

The parameters $a_1$, $a_2$, $a_3$ can be estimated with climb hill algorithm by the test results of day 1, day 2 and day 6 [34]. Then the forgetting curves can be drawn with parameters $a_1$, $a_2$ and $a_3$. And the forgetting status of day 7 can be forecasted.

## Acknowledgements

This paper was supported by Wu Tong Innovation Platform of Beijing Language and Culture University (supported by "the Fundamental Research Funds for the Central Universities") (16PT05)

## Contributions

All authors contributed equally to this work. Yanlu and Yue conducted the experiments, Yanlu and Jinsong analyzed the results. All authors reviewed the manuscript.

## Reference

[1] Ebbinghaus H. (1913) Memory: a contribution to experimental psychology[M]. New York: Columbia University.

[2] Brown, G. D. A., Neath, I., & Chater, N. (2007) A temporal ratio model of memory. Psychological Review, 114, 539–576.

[3] Sverker Sikstrom. (2001) Forgetting curves: implications for connectionist models Cognitive Psychology 45 95–152.

[4] Loftus G. (1985) Evaluating forgetting curves.[J]. Journal of Experimental Psychology: Learning, Memory and Cognition.

[5] Bogartz R. (1990) Evaluating forgetting curves psychologically.[J]. Journal of Experimental Psychology: Learning, Memory and Cognition.

[6] Wixted, J. T., & Ebbesen, E. B. (1991) On the form of forgetting. Psychological Science, 2, 409–415.

[7] Lee Averell, Andrew Heathcote. (2011) The form of the forgetting curve and the fate of memories Journal of Mathematical Psychology 55 25–35

[8] Wixted J, Ebbesen E. (1997) Genuine power curves in forgetting: A quantitative analysis of individual subject forgetting functions[J]. Memory & Cognition.

[9] Jaber M, Bonney M. A. (1997) comparative study of learning curves with forgetting[J]. Applied Mathematical Modelling, 21(8): 523-531.

[10] Fioravanti M, Cesare F. (1992) Forgetting curves in long-term memory: Evidence for a multistage model of retention[J]. Brain and Cognition, 18(2): 116-124.

[11]Simmon, H. A. (1966) A note on Jost's law and exponential forgetting. Psychometrika, 31, 505–506.

[12]M J Murre, Jaap & Dros, Joeri. (2015). Replication and Analysis of Ebbinghaus' Forgetting Curve. PloS one. 10. e0120644. 10.1371/journal.pone.0120644.

[13]Shiffrin, R. M., Lee, M. D., Wagenmakers, E.-J., & Kim, W. J. (2008) A survey of model evaluation approaches with a tutorial on hierarchical Bayesian methods. Cognitive Science, 32, 1248–1284.

[14]Settles, Burr & Meeder, Brendan. (2016). A Trainable Spaced Repetition Model for Language Learning. 1848-1858.

[15]Lee Averell, Andrew Heathcote. (2011) The form of the forgetting curve and the fate of memories Journal of Mathematical Psychology 55 25–35.

[16]Rubin, D.C. (1982) On the retention function for autobiographical memory. Journal of Verbal Learning and Verbal Behavior, 21, 21–38.

[17]Yi Feiyi, Ren Lifeng, Xie Jiaping A Maths Model on Studying and Recalling. (1997) Journal of Mathematical Medicine. 1997, 10 (2): 105-108 in Chinese.

[18]Rubin, David, C.; Hinton, Sean; Wenzel, Amy.(1999) The precise time course of retention. Journal of Experimental Psychology Learning Memory and Cognition 1999

[19]Vlach H, Sandhofer C. (2012) Fast Mapping Across Time: Memory Processes Support Children's Retention of Learned Words[J]. Frontiers in Psychology.

[20]Nembhard D, Uzumeri M. (2000) Experiential learning and forgetting for manual and cognitive tasks[J]. International Journal of Industrial Ergonomics, 25(4): 315-326.

[21]Baddeley, A. D. (1992) Working memory. Science 255, 556–559, 10.1126/science.1736359.

[22]Baddeley, A. D. (2000) The episodic buffer: a new component of working memory? Trends Cogn. Sci. 4, 417–423, 10.1016/S1364-6613(00)01538-2.

[23]Baddeley, A. D. & Logie, R. H. (1999) Working memory: The multiple component model In Models of working memory: Mechanisms of active maintenance and executive control (eds Miyake, A. & Shah, P.) 28–61 (Cambridge University Press, New York).

[24]Gathercole,S.E. A.Baddeley. (1994) Working Memory and Language. Hove,Sussex: Lawrence Erlbaum.

[25]Pina, V., Fuentes, L. J., Castillo, A. & Diamantopoulou, S. (2014) Disentangling the effects of working memory, language, parental education and non-verbal intelligence on children's mathematical abilities. Front. Psychol. 5, 415, 10.3389/fpsyg.2014.00415.

[26]Ericsson K, Kintsch W. (1995) Long-Term Working Memory.[J]. Psychological Review, 102(2): 211-45.

[27]Peter Skehan. (1998) A Cognitive Approach to Language Learning Oxford: Oxford University Press.

[28]Morales J, Calvo A, Bialystok E. (2013) Working memory development in monolingual and bilingual children[J]. Journal of Experimental Child Psychology, 114(2): 187-202.

[29]Alloway T, Gathercole S, Kirkwood H. (2008) Evaluating the Validity of the Automated Working Memory Assessment.[J]. Educational Psychology, 28(7): 725-734.

[30]Curve of Forgetting. http://uwaterloo.ca/counselling-services/curve-forgetting

[31]I. I. Hirschman and D. V. Widder. (1955) The convolution transform Princeton, Princeton University Press.

[32]YUE WANG, Y Zhang. (2007) Neural plasticity in speech acquisition and learning - Bilingualism: Language and Cognition, 10(2) 147-160 2007 - Cambridge Univ Press.

[33]Yue Sun; Jinsong Zhang; Yanlu Xie etc (2013) Perceptual training of Japanse students to discriminate Mandarin Tones 2 and 3   Journal of Tsinghua University(Science and Technology) (Chinese).

[34]Ackley, D.H. (1987) A connectionist machine for genetic hill climbing, Kluwer Academic Pub.